\documentclass[12pt]{article}
\begin{document}

\title{Conceptual Unification of Gravity and Quanta}
\author{Michael Heller\thanks{%
Correspondence address: ul. Powsta\'nc\'ow Warszawy 13/94, 33-110 Tarn\'ow,
Poland. E-mail: mheller@wsd.tarnow.pl} \\
Vatican Observatory, V-00120 Vatican City State \and Leszek Pysiak \\
Department of Mathematics and Information Science, \\
Warsaw University of Technology \\
Plac Politechniki 1, 00-661 Warsaw, Poland \and Wies{\l }aw Sasin \\
Department of Mathematics and Information Science, \\
Warsaw University of Technology \\
Plac Politechniki 1, 00-661 Warsaw, Poland}
\date{\today}
\maketitle

\begin{abstract}
We present a model unifying general relativity and quantum mechanics. The
model is based on the (noncommutative) algebra \mbox{${\cal A}$} \ on the
groupoid $\Gamma = E \times G$ where $E$ is the total space of the frame
bundle over spacetime, and $G$ the Lorentz group. The differential
geometry, based on derivations of \mbox{${\cal A}$}, is constructed. The
eigenvalue equation for the Einstein operator plays the role of the
generalized Einstein's equation. The algebra \mbox{${\cal A}$}, when
suitably represented in a bundle of Hilbert spaces, is a von Neumann algebra
$\mathcal{M}$ of random operators representing the quantum sector of the
model. The Tomita-Takesaki theorem allows us to define the dynamics of
random operators which depends on the state $\varphi $. The same state
defines the noncommutative probability measure (in the sense of Voiculescu's
free probability theory). Moreover, the state $\varphi $ satisfies the
Kubo-Martin-Schwinger (KMS) condition, and can be interpreted as describing
a generalized equilibrium state. By suitably averaging elements of the
algebra \mbox{${\cal A}$}, one recovers the standard geometry of
spacetime. We show that any act of measurement, performed at a given
spacetime point, makes the model to collapse to the standard quantum
mechanics (on the group $G$). As an example we compute the noncommutative
version of the closed Friedman world model. Generalized eigenvalues of the Einstein
operator produce the correct components of the energy-momentum tensor.
Dynamics of random operators does not ``feel'' singularities.
\end{abstract}

\section{INTRODUCTION}
One of the driving forces of scientific progress is the evolution of
concepts, and concepts evolve when they are involved in solving problems.
Currently, the main problem of theoretical physics is to find a sufficiently
rich mathematical structure which, when suitably interpreted, would contain
in itself (as some \textquotedblleft limiting cases\textquotedblright )
physics of gravity and physics of quanta. It is rather obvious that when
this goal is finally reached, it will induce a radical conceptual
revolution. In a series of works (Heller \textit{et al. }1997;
1999; 2004; 2005a,b,c) we have proposed a model, based on nuncommutative
geometry, unifying general relativity and quantum mechanics (with the
perspective of including quantum field theory). We think that the main
attractiveness of this model is its conceptual structure firmly based on its
mathematical architecture. The main idea of the model consists in exploring
a noncommutative algebra \mbox{${\cal A}$}, defined on a transformation
groupoid $\Gamma $ which is given by the action of a group (typically the
Lorentz group) on the frame bundle $(E,\pi _{M},M)$ over spacetime $M$. The
geometry of $M$ (physics of gravity) can be recovered by suitably averaging
elements of \mbox{${\cal A}$}, and quantum sector of the model is obtained
by representing the algebra \mbox{${\cal A}$} \ on a family of Hilbert
spaces associated with the groupoid $\Gamma $. Our approach differs from
that of Connes (1994) and the authors following him (e.g., Chamseddine
\textit{et al.}, 1993; Madore and Mourad, 1998; Madore and Saeger, 1998;
Sitarz, 1994; Moffat, 2000; Chamseddine, 2001; 2004; Szabo, 2006) in that 
that we explore the structure of the groupoid $\Gamma $
and base our construction of the noncommutative differential algebra on 
\mbox{${\cal A}$} \ and the (sub)module of its derivations
[similarly to the approach developed by Dubois-Violette (1988);
see also (Dubois-Violette and Michor, 1994)] whereas Connes
does this on the representation of the corresponding algebra on a
Hilbert space, and he uses differential forms rather than
derivations. We go to the representation of \mbox{${\cal A}$} \
only to recover the quantum sector of our model.

In the present paper, we further develop our model, both its mathematical and
conceptual aspects. We show how strongly these aspects interact with each
other. To make the paper
self-contained, new results (indicated below) are presented in a broader
context of the model's structure. In section 2, we briefly present the
groupoid $\Gamma =E\times G$ (in the present paper $G$ is a noncompact
group) and the noncommutative algebra \mbox{${\cal A}$} \ of smooth compactly supported,
complex valued functions on $\Gamma $ with convolution as multiplication.
The groupoid can be regarded as a space of generalized
symmetries of our model. The differential geometry of the groupoid $\Gamma $ is based on the algebra $
\mathcal{A}$ and its derivations. Derivations are classified into three types: horizontal $V_1$, vertical $V_2$ and inner $V_3$ (Heller \textit{et al.}, 2005c). The pair $(\mathcal{A}, V)$, where $V$ is a subset of the module $\mathrm{Der}(\mathcal{A})$ of all derivations of the algebra $\mathcal{A}$, is called a differential algebra. We introduce it in section 3. The gravitational sector of the model is presented in section 4. It is based on the differential algebra $(\mathcal{A}, V)$ where $V = V_1 \oplus V_2$. We first construct the corresponding differential geometry (connection, curvature, Einstein operator), and then we postulate that the eigenvalue equation for
the Einstein operator should play the role of a generalized Einstein's
equation (no energy-momentum tensor is assumed). This is an important
modification with respect to (Heller \textit{et al.}, 2005c); its best justification being
the result obtained in section 5, where the components of this equation are
computed for the closed Friedman world model. It turns out (by comparison
with the usual Friedman model) that the (generalized) eigenvalues of the Einstein
operator should be interpreted as matter sources. A suitable equation of
state turns out to be encoded in a relationship between different eigenvalues. This
example also shows that the groupoid $\Gamma $ and the noncommutative algebra \mbox{${\cal
A}$} \ are essential elements of the model; without them this result could
not be obtained.
In section 6, we present the quantum sector of the model. The
algebra \mbox{${\cal A}$}, when suitably represented in a bundle of Hilbert
spaces, is a von Neumann algebra $\mathcal{M}$ of random operators (the von
Neumann algebra of the groupoid $\Gamma $). The Tomita-Takesaki theorem,
applied to this algebra, allows us to define the dynamics of random operators which depends on a state $\varphi $ on $\mathcal{M}$. The pair $(\mathcal{M}, \varphi )$ can thus be regarded as a ``dynamic object'' of our model.
In section 7, we summarize the above results by defining a dynamical system for our model. It consists of two equations: the eigenvalue equation for the Einstein operator and the dynamical equation of the Tomita-Takesaki theorem. The first of these equations corresponds to the differential algebra $(\mathcal{A}, V_1 \oplus V_2)$, the second is related, via a suitable representation, to the differential algebra $(\mathcal{A}, V_3)$. In section 8, we discuss some dynamical properties of our model.
If $\varphi $, appearing in the modular evolution, is a faithful and normal state, it also defines the noncommutative
probability measure (Voiculescu, 1985; Voiculescu \textit{et al.}, 1992).
Thus the pair $(\mathcal{M},\varphi )$ is both a \textquotedblleft dynamic
object\textquotedblright\ and a \textquotedblleft probabilistic
object\textquotedblright\ of the model. For the full discussion of these
properties one should refer to (Heller \textit{et al.}, 2005b, 2005c); here
a new element has been added: if the state $\varphi $ satisfies the
Kubo-Martin-Schwinger (KMS) condition, it can be interpreted as describing a
generalized equilibrium state (Rovelli, 1993; Martinetti and Rovelli, 2003).
Therefore, on the fundamental level, dynamics, probability and at least some
thermodynamic properties are encoded in the same mathematical structure.
In section 9, we return to the example of the noncommutative version of the
closed Friedman universe, and explore its quantum sector. The most
intriguing result is that the random dynamics on the fundamental (Planck)
level does not ``feel'' singularities. They emerge, together with spacetime
when the noncommutative regime changes into the usual commutative evolution.
In section 10, we show how to obtain general relativity and quantum mechanics from our model as its limiting cases.
A few remarks, in section 11, concerning perspectives of the model close the paper.

\section{NONCOMMUTATIVE GENERALIZATION OF SPACETIME}

In our first encounter with the special theory of relativity we were not
immediately exposed to the spacetime geometry or to some other abstract
mathematical structures but rather we were instructed how to change from one
inertial reference frame to another inertial reference frame with the help
of a Lorentz transformation. In this sense, the set of pairs of reference
frames and elements of the Lorentz group transforming these frames into one
another forms a natural setting for the special theory of relativity. If a
suitable care is applied, the same procedure could be extended to general
relativity. These considerations lead to the following construction.

Let $\pi _{M}:E\rightarrow M$ be a frame bundle over spacetime $M$ with the
structural group $G=SO_{0}(3,1)$. A fibre $E_{x}=\pi _{M}^{-1}(x)$ over $%
x\in M$ is the set of local reference frames attached to the point $x$. For
any pair of such frames $p,q\in E_{x}$ there exists $g\in G$ such that $p=qg$%
. We see that $G$ acts on $E$, $E\times G\rightarrow E$, along the fibres.
This allows us to construct the Cartesian product
\[
\Gamma =E\times G=\{\gamma =(p,g):p\in E,g\in G\},
\]%
two elements of which, $\gamma _{1}=(p_{1},g_{1})$ and $\gamma
_{2}=(p_{2},g_{2})$ can be composed (multiplied) if $p_{2}=p_{1}g_{1}$ to
give $\gamma _{1}\circ \gamma _{2}=(p_{1},g_{1}g_{2})$. The inverse of $%
\gamma =(p,g)$ is $\gamma ^{-1}=(pg,g^{-1})$. There are two mappings: $%
d(p,g)=p$ and $r(p,g)=pg$, called the \textit{source\/} mapping and the
\textit{target\/} mapping, respectively. The \textit{set of units\/} is
defined to be $\Gamma ^{(0)}=\{(p,e):p\in E\}$ where $e$ is the unit of $G$.
If some natural conditions are satisfied, $\Gamma $ is called \textit{%
groupoid\/} [for definition see (Paterson, 1999; Heller \textit{et al.}
2004; Pysiak 2004)]. If this purely algebraic construction is equipped with
the smoothness structure, it is called a \textit{smooth\/} or \textit{Lie
groupoid\/}.

The above described groupoid $\Gamma $ implements the idea of a space, the
elements of which consist of two reference frames $d(p,g)=p$ and $r(p,g)=pg$%
, and the element $g$ of the Lorentz group $G$ transforming $p$ into $q=pg$.
We refer to $\Gamma $ as to the \textit{transformation groupoid\/}.
The same idea can be implemented by specifying only two reference frames $%
p_{1}$ and $p_{2}$ attached to the same point $x$ of $M$. We thus define
\[
\Gamma _{1}=\{(p_{1},p_{2})\in E\times E:\pi _{M}(p_{1})=\pi
_{M}(p_{2})=x\in M\}
\]%
with the composition law: $(p_{1},p_{2})\circ
(p_{2},p_{3})=(p_{1},p_{3}),\,p_{1},p_{2},p_{3}\in E$. It is a groupoid
called \textit{groupoid of pairs\/}. In fact, $\Gamma $ and $\Gamma _{1}$
are isomorphic as groupoids (Heller \textit{et al.}, 2005c).

As it is well known, the geometry of spacetime $M$ can be
reconstructed in terms of the algebra $C^{\infty }(M)$ of smooth
functions on $M$. Moreover, also Einstein's equations can be
defined in terms of this algebra (Geroch, 1972; Heller, 1992;
Heller and Sasin, 1995a). The natural idea would be to apply the
same strategy to the space $\Gamma $ (or $\Gamma _{1}$). It turns
out that to obtain the case interesting from both mathematical and
physical points of view, we should consider a noncommutative
algebra $\mathcal{A}$ on the groupoid $\Gamma $. A commutative algebra
$(\mathcal{A},\cdot )$ of smooth, complex, compactly supported
functions on $\Gamma $ with the usual pointwise multiplication would give us again a classical geometry.
To obtain a noncommutative algebra we replace the usual pointwise multiplication with the convolution: if
$f_{1},f_{2}\in \mathcal{A}$ then
\[
(f_{1}\ast f_{2})(\gamma )=\int_{\Gamma _{d(\gamma )}}f_{1}(\gamma
_{1})f_{2}(\gamma _{1}^{-1}\gamma )d\gamma _{1}
\]%
where the integration is over all elements $\gamma \in \Gamma $ beginning at
$p=d(\gamma )$ which is denoted by $\Gamma _{d(\gamma )}$. Let us notice that convolution
algebras play the crucial role in harmonic analysis and in representation theory, both in
in the case of groups and groupoids. The algebra $(
\mathcal{A},\ast )$, being now noncommutative, is nonlocal. It has no
maximal ideals which would correspond to points and their neighbourhoods in $%
\Gamma $. The groupoid $\Gamma $ is replaced by its noncommutative version,
i.e., by the \textquotedblleft virtual space\textquotedblright\
corresponding to the algebra $(\mathcal{A},\ast )$. Keeping this in mind we
shall denote this algebra by $\mathcal{A}=(C_{c}^{\infty }(\Gamma ,\mathbf{C}%
),\ast )$. As we shall see in the following sections, this algebra is rich
enough to contain it itself both relativistic and quantum structures.

If we chose the groupoid $\Gamma _{1}$ rather than the groupoid $\Gamma $,
we should define the algebra $\mathcal{A}_{1}$ on $\Gamma _{1}$ via the
isomorphism $J:\mathcal{A}_{1}\rightarrow \mathcal{A}$ given by $J(f)(\gamma
)=f(p,pg)$ for $f\in \mathcal{A}_{1}$ and $\gamma =(p,g)$ (Heller \textit{et
al.} 2005c).

\section{DIFFERENTIAL ALGEBRA}

The first thing we must ensure is that the noncommutative geometry based on
the algebra $\mathcal{A}$ should allow us to recover the usual
(noncommutative) spacetime geometry as a special case. The natural way of
doing this would be by restricting the algebra $\mathcal{A}$ to its center $%
\mathcal{Z}(\mathcal{A})$ (i.e., to the subset of $\mathcal{A}$
consisting of all these elements that commute with all elements of
$\mathcal{A}$). Unfortunately, $\mathcal{Z}(\mathcal{A})=\{0\}$.
It turns out, however, that the lifting of the algebra $C^{\infty
}(M)$ to the total space $E$ of the frame bundle over $M$, i.e.,
the set $Z = \pi_M^*(C^{\infty }(M))$ (which is, of course
isomorphic with $C^{\infty }(M)$), can be regarded as an ``outer
center'' of the algebra $\mathcal{A}$. To be more precise,
although the functions belonging to $Z$ are not compactly
supported, we can define their action on the algebra $\mathcal{A}$, $\alpha : Z
\times \mathcal{A} \rightarrow \mathcal{A}$, by
\[
\alpha(f,a)(p,g) = f(p)a(p,g),
\]
$f \in Z, a \in \mathcal{A}$. We say that the algebra $\mathcal{A}$ is a
module over $Z = \pi_M^*(C^{\infty }(M))$. This fact allows us to develop a
noncommutative geometry based on the algebra $\mathcal{A}$ which will be a
true generalization of the usual spacetime geometry.

One can base the noncommutative geometry either on differential forms
defined in terms of the algebra $\mathcal{A}$ (Connes, 1994;
Chamseddine \textit{et al.}, 1993; Madore and Mourad, 1998; Madore
and Saeger, 1998; Sitarz, 1994), or in terms of derivations of this algebra
(Dubois-Violette, 1988; Dubois-Violette and Michor, 1994; Madore, 1999).
The first method is more common, but the second method is closer to the
usual way of doing (commutative) differential geometry. In our case, there
is plenty of derivations, and the second method seems more appropriate.

\ A \emph{derivation\/} of the algebra $\mathcal{A}$ is a linear map $v:%
\mathcal{A}\rightarrow \mathcal{A}$ satisfying the Leibniz rule
\[
v(a,b)=v(a)b+av(b) .
\]
It can be thought of as a generalization of the vector field concept. The
set of all derivations of the algebra $\mathcal{A}$ will be denoted by $%
\mathrm{Der}(\mathcal{A} )$. It has the algebraic structure of a Z-module.

Let $\bar{X}$ be a vector field on $E$; we shall write $\bar{X}\in \mathcal{X%
}(E)$. Let us also assume that $\bar{X}$ is a \emph{right invariant\/}
vector field, i.e.
\[
(\mathcal{R}_{g})_{\ast p}\bar{X}(p)=\bar{X}(pg)
\]%
for every $g\in G$. The \emph{lifting\/} of $\bar{X}$ to $\Gamma $ is
defined to be
\[
\bar{\bar{X}}(p,g)=(\iota _{g})_{\ast p}\bar{X}
\]%
where the inclusion $\iota _{g}:E\hookrightarrow E\times G$ is given by $%
\iota _{g}(p)=(p,g)$. It can be shown that the lifting of a right invariant
vector field $\bar{X}\in \mathcal{X}(E)$ to $\Gamma $ is a derivation of the
algebra $\mathcal{A}$ (Heller \textit{et al.}, 2005c).

Let $\bar {X}\in \mathcal{X} (E)$ be a right invariant vector field. If it
satisfies the condition $(\pi_M)_*\bar { X}=0$ it is said to be a \emph{%
vertical\/} vector field. Such vector fields, when lifted to $\Gamma$, are
derivations of the algebra $\mathcal{A}$ and are called \emph{vertical
derivations}.

Let us suppose that a connection is given in the frame bundle $\pi
_{M}:E\rightarrow M$ [for details see (Heller \textit{et al.}, 2005c)]. With
the help of this connection we lift a vector field $X$ on $M$ to $E$, i.e., $\bar{X%
}(p)=\sigma (X(\pi _{M}(p)),\,\pi _{M}(p)=x\in M$ where $\sigma $ is a lifting
homomorphism. This vector field is
right invariant on $E$. If we lift it further to $\Gamma $
\[
\bar{\bar{X}}(p,g)=(\iota _{g})_{\ast p}\bar{X}(p)\in \mathcal{X}(\Gamma ),
\]%
it preserves its right invariance property, and is a derivation of the
algebra $\mathcal{A}$. We call it a \emph{horizontal derivation\/} of $%
\mathcal{A}$.

The algebra $\mathcal{A}$ has also derivations typical for noncommutative
algebras. They are called \emph{inner derivations}, denoted by $\mathrm{Inn}(%
\mathcal{A})$, and defined to be
\[
\mathrm{Inn}(\mathcal{A})=\{ad(a):a\in \mathcal{A}\}
\]%
where $(ad(a))(b):=a\ast b-b\ast a$. Of course, for commutative algebras all
such derivations vanish. It is important to notice that the mapping $\Phi
(a)=ad(a)$, for every $a\in \mathcal{A}$, establishes the isomorphism between
the algebra $\mathcal{A}$ and the space $\mathrm{Inn}(\mathcal{A})$ as $Z$%
-moduli (Heller \textit{et al.}, 2004).

By the \emph{differential algebra\/} we understand a pair $(\mathcal{A},V)$
where $\mathcal{A}$ is (not necessarily commutative) algebra and $V\subset
\mathrm{Der}(\mathcal{A})$ is a (sub)module of its derivations. In section 4, we construct the gravitational sector of our model basing on the differential algebra ($\mathcal{A},V)$
where $V=V_{1}\oplus V_{2}$ with $V_{1}$ and
$V_{2}$ being horizontal and vertical derivations, respectively. The $Z$-module $V_3 = \mathrm{Inn}(\mathcal{A})$ is responsible for quantum effects; it is taken into account in section 6.

\section{GRAVITATIONAL SECTOR}

\subsection{Geometry}

In the present
section, we compute the \textquotedblleft groupoid
geometry\textquotedblright\ for the case when $V = V_1 \oplus V_2$ and $G$ is a noncompact
and semisimple group (which includes the group $SO_{0}(3,1)$); the case with a
finite group $G$ was treated in (Heller \textit{et al.}, 2005a).
As the metric $\mathcal{G}:V\times V \rightarrow Z$ we choose
\[
\mathcal{G}(u,v)=\bar {
g}(u_1,v_1)+\bar {k} (u_2,v_2)
\]
where $u_1,v_{_1}\in V_1,u_2,v_2\in V_2$. The metric $\bar { g}$ is evidently the lifting of the
metric $g$ on spacetime $M$, i.e.,
\[
\bar {g}(u_1,v_1)= \pi_M^{*}(g(X,Y))
\]
where $X,Y \in \mathcal{X} (M).$ We assume that the metric $ \bar {k}$ is of the Killing type. In principle, more general metrics than $\mathcal{%
G}$ could also be considered.

The next step in our construction is to define preconnection by the Koszul
formula
\[
(\nabla _{u}^{\ast }v)w=\frac{1}{2}[u(\mathcal{G}(v,w))+v(\mathcal{G}%
(u,w))-w(\mathcal{G}(u,v))
\]%
\[
+\mathcal{G}(w,[u,v])+\mathcal{G}(v,[w,u])-\mathcal{G}(u,[v,w]).
\]%
In (Heller \textit{et al.}, 2005c) we have proved that if $V$ is a $Z$%
-module of derivations of an algebra $(\mathcal{A},\ast )$ such that $V(Z)=\{0\}$ then, for
every symmetric nondegenerate tensor $g:V\times V\rightarrow Z$, there
exists exactly one connection $g$-consistent with the preconnection $\nabla
^{\ast }$ which is given by
\[
\nabla _{u}v=\frac{1}{2}[u,v].
\]%
We assume that, for $V_{2}$, the metric is of the Killing form
\[
g(u,v)=\mathrm{Tr}(u\circ v)
\]%
(the $g$-consistency condition is clearly satisfied). For the group $ G$
(which in our case is semisimple) the Killing form is
\[
\mathcal{B}(V,W)=\mathrm{Tr}(a d(V)\circ ad(W))
\]
where $V,W$ are elements of the Lie algebra $\underline g$ of the group $G$.
The Killing form $\mathcal{B}$ is nondegenerate. Since the tangent space to
any fiber $E_x,x\in M$ at a fixed point $p \in E$, is isomorphic to $\underline g$, each invariant vertical
vector field $\bar{X}$ can be represented by a $\underline g$-valued
function on $E$, and one can prove that $\mathcal{B}(\bar X(p), \bar Y(p))$
depends only on $\pi_M(p) \in M$. Therefore, the metric $\bar {
k}%
:V_2\times V_2\rightarrow Z$ is given by
\[
\bar {k}(\bar{\bar {
X}},\bar{\bar {Y}})= \mathcal{B}(\bar X(p), \bar Y(p)).
\]

The trace for the algebra $\mathcal{A}_1$ (which is isomorphic to the
algebra $\mathcal{A}$) is given by
\[
\mathrm{Tr}(a)(p)=\int_G a(pg , pg)dg.
\]
It is clear that $\mathrm{Tr}(a) \in Z$.

The \emph{curvature}, for all  $V_i,\,i=1,2$%
, is defined in the usual way
\[
\stackrel{i}{R}(u,v)w =\stackrel{i}{\nabla}_ u\stackrel{i}{\nabla}_ v w -%
\stackrel{i}{\nabla}_ v\stackrel{i}{\nabla}_ u w -\stackrel{i}{\nabla}_{
[u,v]}w.
\]

For $i=2$, we readily compute
\[
\stackrel{2}{R}(u,v)w =-\frac 14[[u,v],w].
\]

For $i=1,2$ and every endomorphism $T:V_i\rightarrow V_i$, there exists the
usual trace $\mathrm{Tr}(T)\in Z$, and we can define $\stackrel{i}{R}%
_{uw}:V_i\rightarrow V_ i$ by
\[
\stackrel{i}{R}_{uw}( v)=\stackrel{i}{R}(u,v) w.
\]
Consequently, the \emph{Ricci curvature} is
\[
\stackrel{i}{\mathbf{r} \mathbf{i}\mathbf{c}}(u,w) =\mathrm{Tr}(\stackrel{i}{R}%
_{uw} ),
\]
and the \textit{adjoint Ricci operator} $\stackrel{i}{\mathcal{R}}: V_i
\rightarrow V_i$ is given by
\[
\stackrel{i}{\mathbf{r} \mathbf{i}\mathbf{c}}(u,w) =\stackrel{i}{\mathcal{G}} (%
\stackrel{i}{\mathcal{R}} (u),w)
\]
where $\stackrel{1}{\mathcal{G}} =\bar {g}$ and $\stackrel{2}{\mathcal{G}}=%
\bar {k}$. If the metric $\stackrel{i}{\mathcal{G}}$ is nondegenerate, there
exists the unique $\stackrel{i}{\mathcal{R}}$ satisfying the above equation
for every $w\in V_i$.

The \emph{curvature scalar\/} is
\[
\stackrel{i}{r}=T r(\stackrel{i}{\mathcal{R}} ).
\]

For $V_2$ (for which the usual trace exists) we compute
\[
\stackrel{2}{\mathbf{r} \mathbf{i}\mathbf{c}}(u,w)=\frac 14\bar {k}(u,w)
\]
for every $u,w\in V_ 2$.

\subsection{Generalized Einstein Equation}
In the present paper we postulate that the \emph{generalized Einstein
equation\/} should have the form of the eigenvalue equation for the Einstein
operator $\mathbf{G} := \mathcal{R} - \frac{1}{2}r \mathrm{id}_V$ where $r =
\mathrm{Tr}\mathcal{R}$. Let us notice that we do not assume \textit{a priori} energy-momentum tensor in any form. The
motivation for this assumption is that on the fundamental level we
expect to have a ``pure pregeometry'', and the ``matter content'' should be
somehow produced from it at a later stage. As we shall see in the next
section, this is indeed the case, at least for the noncommutative version of
the closed Friedman world model.

The eigenvalue equation for the Einstein operator is
\begin{equation}
{\mathbf{G}}-\tau \mathrm{id}_{V}=0  \label{Einstein}
\end{equation}%
where $\tau \in Z$ and $v\in V$; $\tau $ will be called a generalized eigenvalue of the operator $\mathbf{G}$ (generalized, because it is a function rather than a number).

If $v\in V_1$, equation (\ref{Einstein}) reflects essentially the geometry of space-time $M$.

If we assume that the metric $\bar{k}$ is of the Killing type then for $v\in V_2$ and $G=SO_0(3,1)$ the dimension of
the $Z$-module $V_2$ is equal to the dimension of the Lie algebra of $G$
which is 6-dimensional. Consequently, $\mathrm{Tr}(\mathrm{id}_{V_2})=6$.
Since, in this case, $\mathcal{R} = \frac{1}{4}\mathrm{id}_{V_2}$, we have $%
r=\frac{3}{2}$, and the Einstein equation assumes the form
\[
(\tau + \frac{1}{2})(v)=0.
\]
We see that for $\tau = -\frac{1}{2}$ every $v \in V_2$
solves this equation, and for $\tau \neq -\frac{1}{2}$ there is only the
trivial solution.

\section{NONCOMMUTATIVE CLOSED FRIEDMAN UNIVERSE}
As a simple example let us consider the closed Friedman world model. Its spacetime $M=\{(\eta
,\chi ,\theta ,\varphi ):\eta\in (0,T),(\chi ,\theta ,\varphi)\in
S^3\}=(0,T)\times S^3,$ where $(0,T)\subset \mathbf{R}$, carries the metric
\[
ds^2=R^2(\eta )(-d\eta^2+d\chi^2+\sin^2\chi (d\theta^ 2+\sin^2\theta
d\varphi^2)).
\]
The initial singularity is characterized by: $R^2(\eta )\rightarrow 0$ as $%
\eta\rightarrow 0$, and the final singularity by: $R^2(\eta )\rightarrow 0$
as $\eta\rightarrow T$.

Let $(\pi_M: E \rightarrow M)$ be a frame bundle over $M$ where
\[
E=\{(\eta ,\chi ,\theta ,\varphi,\lambda ):(\eta ,\chi ,\theta ,\varphi )\in
M,\lambda\in \mathbf{R}\}=M\times \mathbf{R}.
\]
The structural group of the frame bundle is
\[
G=\{\left(%
\begin{array}{cccc}
\cosh t & \sinh t & 0 & 0 \\
\sinh t & \cosh t & 0 & 0 \\
0 &  0 &  1 &  0 \\
0 &  0 &  0 &  1 \\
\end{array}
\right),\, t\in \mathbf{R}\},
\]
To have the orthonormal frames we make the transformation $\partial_{
\mu}\rightarrow\frac 1{R(\eta )}\partial_{\mu}$. This group of ``Lorentz rotations'' (Carmeli, 1977, p. 22) has been chosen by us because, in spite of its simplicity, it gives an insight into many aspects of the general case [see (Dodson, 1978, sec. 3)].
\par
The space of the pair
groupoid is given by
\[
\Gamma =\{(\eta ,\chi ,\theta ,\varphi ,\lambda_1,\lambda_
2):\lambda_1,\lambda_2\in \mathbf{R}\}.
\]
If $a,b\in \mathcal{A}=C_c^{\infty}(\Gamma ,\mathbf{C})$ then
\[
(a*b)(\eta ,\chi ,\theta ,\varphi ,\lambda_1,\lambda_ 2)=\int_{\mathbf{R}%
}a(\eta ,\chi ,\theta ,\varphi ,\lambda_ 1,\lambda )b(\eta ,\chi ,\theta
,\varphi ,\lambda ,\lambda_ 2)d\lambda .
\]
The ``outer center'' of this algebra is $Z=\{a(\eta ,\chi ,\theta , \varphi):(\eta ,\chi ,\theta ,\varphi )\in
M\}.$ Since the convolution is defined on the groupoid rather than on the
group, the algebra $\mathcal{A}$ is noncommutative in spite of the fact that
the group $G=\mathbf{R}$ is Abelian.

We consider the $Z$-submodule $ V=V_1\oplus V_2$ of
horizontal derivations and vertical derivations of the algebra $\mathcal{A}$%
. The metric on $V$ is
\begin{eqnarray*}
ds^2 &  = &  -R^2(\eta )d\eta^2+ \mbox{}R^2(\eta )d\chi^2+R^2(\eta )\sin^
2(\chi )d\theta^2+ \\
&  &  R^2(\eta )\sin^2(\chi )\sin^2(\theta )d\varphi^2+d\lambda^2.
\end{eqnarray*}

The Einstein operator is of the form $\mathbf{G} = G^c_{\phantom{c}d}=%
\mathrm{diag}\{B,h,h,h,q\}$ where $B = -3\frac 1{R^2(t)}-3\frac {%
R^{\prime}{}^2(t)}{R^4(t)}$, $h=\displaystyle-\frac 1{R^2(t)}+\frac {%
R^{\prime}{}^2(t)}{R^4( t)}-2\frac {R^{\prime\prime}(t)}{R^3(t)}$ and $q=%
\displaystyle-3\frac 1{R^2(t)}-3\frac {R^{\prime\prime} (t)}{R^3(t)}$.
We assume the field equation in the form of the eigenvalue equation for
the Einstein operator $\textbf{G}: V \rightarrow V$, i.e.,
\begin{equation}
\textbf{G}(u)=\tau \cdot u,
\label{EinsteinF}
\end{equation}
or in the matrix form

\[
\left(\begin{array}{ccccc}
B& 0& 0& 0& 0\\
0& h& 0& 0& 0\\
0& 0& h& 0& 0\\
0& 0& 0& h& 0\\
0& 0& 0& 0& q\end{array}
\right)\left(\begin{array}{c}
u_1\\
u_2\\
u_3\\
u_4\\
u_5\end{array}
\right)=\tau\left(\begin{array}{c}
u_1\\
u_2\\
u_3\\
u_4\\
u_5\end{array}
\right)
\]
where $u_1, u_2,u_3,u_4 \in V_1, \,u_5 \in V_2$.

Here $\tau = (\tau_{1}, \ldots ,\tau_5)$, where $\tau_i = 1, 2, \ldots ,5$, are generalized eigenvalues of $\textbf{G}$. We find them by solving the equation
\begin{equation}
{\rm det}(\textbf{G}-\tau \cdot \textbf{I}) = 0.
\end{equation}
The solutions are
\begin{equation}
\tau_B = -3\frac 1{R^2(\eta)}-3\frac {R'{}^2(\eta)}{R^4(\eta)},
\label{B}
\end{equation}
\begin{equation}
\tau_h=\displaystyle-\frac 1{R^2(\eta)}+\frac {R'{}^2(\eta)}{R^4(
\eta)}-2\frac {R^{\prime\prime}(\eta)}{R^3(\eta)},
\label{h}
\end{equation}
\begin{equation}
\tau_q=\displaystyle-3\frac 1{R^2(\eta)}-3\frac {R^{\prime\prime}
(\eta)}{R^3(\eta)}.\label{q}
\end{equation}

The eigenvectors corresponding to these generalized eigenvalues $\tau_i $ form the submodules $W_i$, $i = 1, \ldots ,5$ of $V$
\[
V_1 \oplus V_2 = W_1 \oplus W_2\oplus W_3 \oplus W_4 \oplus W_5,
\]
or
\[
V_1 \oplus V_2 = W_B \oplus W_h\oplus W_q
\]
where $W_B = W_1$ is a 1-dimensional submodule corresponding to the generalized eigenvalue $\tau_B$, $W_h=   W_2\oplus W_3 \oplus W_4$ is a 3-dimensional submodule corresponding to the generalized eigenvalue $\tau_h$, and $W_q = W_5$ is a 1-dimensional submodule corresponding to the generalized eigenvalue $\tau_q$.

Let us notice that $\textbf{G}$ is a homothety on $W_i$ with the factor $\tau_i$.

By comparing equations (\ref{B}) and (\ref{h}) with the components of the perfect fluid energy-momentum tensor for the Friedman world model, we easily identify
\[
\tau_B = 8\pi G\rho(\eta),\]
\[
\tau_h = -8\pi G p(\eta)
\]
where $G$ is the Newtonian gravitational constant, $\rho $ and $p$ are density and stress functions, respectively, and the velocity of light $c=1$. If we denote
\[
T^0_0 = \frac{\tau_B}{8\pi G},
\] \[
T^i_k = - \frac{\tau_h}{8\pi G}\delta^i_k, \ i,k = 1,2,3,
\]
we obtain the components of the energy-momentum tensor
$T^{\mu }_{\nu },\:\mu , \nu = 0,1,2,3$ as generalized eigenvalues
of the Einstein operator $\textbf{G}$ corresponding to the
submodules $W_B$ and $W_h$, respectively.

And what about equation (\ref{q})? We easily verify that
\begin{equation}
\tau_q = 4\pi G(\rho (\eta ) -3p(\eta ))
\label{eqstate}
\end{equation}
which leads to
\begin{equation}
\tau_q = \frac{1}{2}\tau_B + \frac{3}{2}\tau_h.
\label{stateeq}
\end{equation}
\par
It can be easily seen that $\tau_q$ is the trace of $T^{\mu }_{\nu}$. Let us also notice that equation (\ref{eqstate}) is related to the equation of state for the Friedman model. Indeed,
\begin{itemize}
\item
if $\tau_q = 4\pi G\rho $ then we have the equation of state for dust $p = 0$;
\item
if $\tau_q = 0$ then we have the equation of state for radiation $p = (1/3)\rho $;
\item
if $\tau_q = -16 \pi Gp$ then we have Zeldovitch's stiff equation of state $p = - \rho $.
\end{itemize}
As we can see, the remaining generalized eigenvalue is responsible for the equation of state. From the mathematical point of view, any formula satisfying (\ref{stateeq}) can serve as an equation of state, although only some of them have a physical meaning.
\par
The usual Einstein equations are obtained in a straightforward way
\[
\textbf{G}|_{W_B \oplus W_h} = 8 \pi G \cdot T,
\]
where $T = T^{\mu }_{\nu }$, which read
\begin{equation}
8\pi G\rho(\eta) = -3\frac 1{R^2(\eta)}-3\frac {R'{}^2(\eta)}{R^4(\eta)},
\label{density}
\end{equation}
\begin{equation}
8\pi Gp(\eta) = \frac 1{R^2(\eta)}-\frac {R'{}^2(\eta)}{R^4(
\eta)}+2\frac {R^{\prime\prime}(\eta)}{R^3(\eta)}.
\label{pressure}
\end{equation}
\par
It is straightforward to verify that solving equations (\ref{density}) and (\ref{pressure}) for the above equations of state, we obtain the well known solutions
\begin{itemize}
\item
$R(\eta ) =\frac{4\pi m}{3}(1- \cos (\eta ))$, where $m = \rho (\eta )R^3(\eta ) = {\rm const}$, for dust;
\item
$R(\eta ) = \sqrt{\frac{8 \pi G \textsl{M}}{3}}\sin (\eta )$, where $\textsl{M}=\rho (\eta )R^4(\eta )= {\rm const}$, for radiation; and
\item
$R(\eta ) = 2G^{1/4}(\frac{2\pi}{3})^{1/4}\textsl{N}^{1/4}\sqrt{\frac{\tan (\eta }{1+\tan^2 (\eta )}}$, where $\textsl{N}= \rho (\eta )R^6(\eta ) = {\rm const }$, for Zeldovitch's stiff matter.
\end{itemize}
\par

We have started with field equation (\ref{EinsteinF}) understood
as the eigenvalue equation for the Einstein operator, and by
solving the eigenvalue problem we were able to produce the perfect
fluid energy momentum tensor. No matter source has been assumed
{\it a priori}. In this sense, we can say that in the
noncommutative closed Friedman model geometry generates matter. It
was an old Wheeler's idea to produce ``matter out of pregeometry''
(Wheeler, 1980); the latter being ``a combination of hope and
need, of philosophy and physics and mathematics and logic''
(Misner \textit{et al.} 1973, p. 1203). The effect presented above
can be regarded as a step towards the implementation of
this idea in the context of a concrete mathematical model.
\par
In section \ref{Frieddyn} we discuss some aspects of the quantum sector of the closed Friedman model.

\section{QUANTUM SECTOR}
\subsection{Algebra of Random Operators \label{subs41}}
The quantum sector of our model can be extracted from the groupoid algebra \mbox{${\cal
A}$} \ with the help of its regular representation in the Hilbert space $%
\mathcal{H}^p=L^2(\Gamma^ p)$, for every $p\in E$ ($\Gamma^ p$ being the set of all 
elements of $\Gamma$ ending at $p$),
\[
\pi_p:\mathcal{A}\rightarrow \mathcal{B}(\mathcal{H}^p) ,
\]
where $\mathcal{B}(\mathcal{H}^ p)$ is the algebra of bounded operators on
the Hilbert space $\mathcal{H}^p$. The representation $\pi_p$ is given by
\[
(\pi_p(a)\psi )(\gamma )=\int_{\Gamma_{d(\gamma )}}a(\gamma_1)\psi (
\gamma_1^{-1}\circ\gamma )d\gamma_1
\]
where $a\in \mathcal{A} ,\psi\in \mathcal{H}^p, \gamma ,\gamma_1\in\Gamma$.
Here the Haar measure on the group $G$, transferred to each fiber of $ \Gamma
$, forms a Haar system on $\Gamma$.

It is interesting to notice that the quantum sector of our model
exhibits strong probabilistic properties from the very beginning
(without putting them by hand into the model). We shall show that
every $a\in \mathcal{A}$ generates a random operator $
r_a=(\pi_p(a))_{p\in E}$, acting on a collection of Hilbert
spaces $ \{\mathcal{H}^p\}_{p \in E}$ where $\mathcal{H}^
p=L^2(\Gamma^p)$.

An operator $r_{a}$ is a \emph{random operator\/} if it satisfies the
following conditions (Connes, 1994).

(1) If $\xi_p,\eta_p \in \mathcal{H}^p$ then the function $ E\rightarrow
\mathbf{C}$, given by
\[
E\ni p\mapsto (r_a \xi_p,\eta_p),
\]
$a\in \mathcal{A}$, is measurable in the usual sense (i.e., with respect to
the usual manifold measure on $E$).

(2) The operator $r_ a$ is \emph{bounded}, i.e., $ ||r_a|| < \infty$ where
\[
||r_a||=\,\mathrm{e}\mathrm{s} \mathrm{s}\,\mathrm{s}\mathrm{u} \mathrm{p}%
||\pi_p(a)||.
\]
Here ``ess sup'' denotes essential supremum, i.e., supremum modulo zero
measure sets.

\ In our case, both these conditions are satisfied. Let us also notice that $%
\pi_p(a)$, for every $p\in E$, is a bounded operator on $\mathcal{H}^p$.

\ There exist the isomorphisms $I_p: L^2(G) \rightarrow \mathcal{H}^p$, for
every $p \in E$, given by
\[
(I_p\psi )(pg^{-1},g) = \psi (g)
\]
which can be used to establish the relationship between random operators and
operators on $\mathcal{H}^p$. These isomorphisms will play an important role
in our further analysis.

Let us denote by $\mathcal{M}_{0}$ the algebra of equivalence classes
(modulo equality almost everywhere) of bounded random operators $r_{a},a\in
\mathcal{A}$, and let us define $\mathcal{M}=\mathcal{M}_{0}^{\prime \prime }
$ where $\mathcal{M}_{0}^{\prime \prime }$ denotes the double commutant of $%
\mathcal{M}_{0}$. The algebra $\mathcal{M}$ is a von Neumann algebra
(Connes, 1994). We shall call it the \emph{von Neumann algebra of the
groupoid} $\Gamma $.

As well known, the work of Segal, Kastler, Haag, Gelfand an others has
developed an algebraic description of quantum systems (with both finite and
infinite number of degrees of freedom). It consists of the following main
ingredients: (1) an abstract $C^{\ast }$-algebra encoding, among others,
observables of the system and its statistical properties, (2) automorphisms
of this algebra encoding the dynamics of the system and its symmetries, and
(3) a state functional defining a probability measure on observables (Alicki
and Fannes, 2005). It can be shown that $\mathcal{M}$ itself is a $C^{\ast }$%
-algebra (Murphy, 1990) and, as we shall see in the following subsections,
it satisfies all the above requirements. Therefore, the algebra $\mathcal{M}$
is a mathematical structure that can be interpreted as a true generalization
of the usual quantum theory in its algebraic formulation.

\subsection{Noncommutative Dynamics}

The mathematical structure of our model is encoded in the differential algebra $(\mathcal{A}, \mathrm{Der}(\mathcal{A}))$. As we have seen in section 4, the field equation of the gravitational sector was obtained by considering the $Z$-submodule $V = V_1 \oplus V_2 \subset \mathrm{Der}(\mathcal{A})$; in the present subsection we consider the $Z$-submodule $V_3 = \mathrm{Inn}(\mathcal{A})$, and show that this leads to the dynamic equation of the quantum sector of our model.

Let us then consider the differential algebra $(\mathcal{A}, \mathrm{Inn}(\mathcal{A}))$, and let us remember that $\mathcal{A}$ and $\mathrm{Inn}(\mathcal{A})$ are isomorphic as $Z$-moduli (the isomorphism id given by $a \mapsto \mathrm{ad}_a$, see above section 3). Moreover, every $a \in \mathcal{A}$ generates the random operator $r_a = (\pi_p(a))_{p \in E}$ and, as we have seen, the space $\mathcal{M}_0$ of such operators can be completed to the von Neumann algebra $\mathcal{M}$. [It is interesting to notice that all derivations of any von Neumann algebra are inner (Dixmier, 1981, pp.349-357)]. In (Heller \textit{et al.}, 2005c) we have shown that the Tomita-Takesaki
theorem can be applied to the algebra $\mathcal{M}$ (this algebra is
semifinite) to obtain the evolution of random operators [see also (Connes
and Rovelli, 1994)]. Let us define the Hamiltonian as $H_{p}^{\varphi }=%
\mathrm{L}\mathrm{o}\mathrm{g}\hat{\rho}_{p}^{\varphi }$, where $\hat{\rho}%
(p)$ is a positive, trace class operator in $\mathcal{B}(\mathcal{H}^{p})$,
and $\varphi $ is a state on $\mathcal{M}$ defined to be
\[
\varphi (A)=\int_{M}tr(\hat{\rho}(p)A(p))d\mu (x).
\]%
Let us notice that the integrated function depends only on $x\in M$. We
additionally assume that $\varphi (1)=1$ (Heller \textit{et al.}, 2005c).
On the strength of the Tomita-Takesaki theorem there exists a
one-parameter group of automorphisms $\sigma _{t}^{\varphi }$, called \emph{%
modular group},
\begin{equation}
\sigma _{t}^{\varphi }(r_{a}(p))=e^{itH_{p}^{\varphi
}}r_{a}(p)e^{-itH_{p}^{\varphi }}  \label{TT}
\end{equation}%
for every $p\in E$. Equation (\ref{TT}) can also be written in the form
\begin{equation}
i\hbar \frac{d}{dt}\sigma _{t}^{\varphi }(r_{a}(p))=[r_{a}(p),H_{p}^{\varphi
_{{}}}].  \label{TT2}
\end{equation}
where the Planck constant $\hbar $ is inserted to have the
correspondence with quantum theory. This equation describes the
state dependent evolution of random operators with respect to the
parameter $t\in \mathbf{R}$ of the modular group. We can say that
the pair $(\mathcal{M},\varphi )$ is a \emph{dynamical object\/}
of our model. Equation (\ref{TT2}) is a generalization of the
Heisenberg equation of the usual quantum mechanics with the only
difference that it now depends on the state $\varphi $. There
exists the canonical way of getting rid of this dependence based
on the following Connes-Nicodym-Radon construction (Sunder,
1987).  Let $\mathcal{U}=\{u\in \mathcal{M}:uu^{\ast }=u^{\ast
}u=1\}$ be the unitary group of the algebra $\mathcal{M}$. Two
automorphisms $\alpha _{1}$ and $\alpha _{2}$ of the von Neumann algebra $%
\mathcal{M}$ are said to be \textit{inner equivalent} if there is an element
$u\in \mathcal{U}$ such that
\[
u\alpha _{2}(r)=\alpha _{1}(r)u
\]
for $r\in \mathcal{M}$. The set of equivalence classes of this relation is
called the group of \textit{outer automorphisms\/} denoted by $Out(\mathcal{M%
})$. In general, the modular transformations $\sigma _{t}^{\varphi }$ are
not inner automorphisms of $\mathcal{M}$, but they canonically project onto
the same one-parameter group in $Out(\mathcal{M})$ which is independent of
the state $\varphi $. However, we have demonstrated in (Heller \textit{et al.%
}, 2005b) that the von Neumann algebra $\mathcal{M}$ of our model is
semifinite, and the Dixmier-Takesaki theorem (Connes, 1994) states that if $%
\mathcal{M}$ is semifinite then every state dependent modular evolution is
inner equivalent to the trivial one. This means that the state independent
\textquotedblleft outer evolution\textquotedblright\ is trivial: there is a
state independent time but it does not flow (or nothing happens in it). This
once more demonstrates the radical character of the noncommutative regime of
our model (in its present form) -- it admits only a state dependent
dynamics. In the following subsection, we shall show how this peculiarity is
related to the concept of probability.
\par
To sum up. There is an isomorphism $\mathrm{Inn}(\mathcal{A}) \rightarrow \mathcal{M}_0$ (by $\mathrm{ad}(a) \mapsto a \mapsto r_a$, see the next section), and the latter space can be completed to the von Neumann algebra which, together with a suitable state $\varphi $, forms a dynamical object of our model. We may thus say that equations (\ref{TT}) or (\ref{TT2}) are natural dynamical equations for the quantum sector of our model, and that they can be traced back to the differential algebra $(\mathcal{A}, \mathrm{Inn}(\mathcal{A}))$.
\par
\section{DYNAMICAL SYSTEM}
Mathematical structure of our model is encoded in the differential algebra $(\mathcal{A}, \mathrm{Der}(\mathcal{A}))$ where $\mathrm{Der}(\mathcal{A}) = V_1 \oplus V_2 \oplus V_3$. The $Z$-submodule $V = V_1 \oplus V_2$ is responsible for the gravitational sector, and the field equation for this sector assumes the form of the generalized eigenvalue equation (\ref{Einstein}). The submodule $V_3$ is responsible for the quantum sector, and the dynamics of this sector is given by ``modular equation'' (\ref{TT}) [resp. (\ref{TT2})]. Therefore, we can say that equation (\ref{Einstein})
 and (\ref{TT}) [resp. (\ref{TT2})] form the ``dynamical system'' of our model. However, these two equations are of a very different character: equation (\ref{Einstein}) is classically geometric, and equation (\ref{TT}) is quantum probabilistic. Is there a possibility to make these equations ``more unified''? Indeed, there is such a possibility. To show it, we must first prove the following lemma.
\par
\textbf{Lemma.} The mapping $\pi : \mathcal{A} \rightarrow \mathcal{M}_0$, given by $\pi (a) = (\pi_p(a))_{p \in E}$, is an isomorphism of algebras.

\par
\textbf{Proof:} The mapping $\pi $ is an isomorphism since, for every $p \in E$, $\pi_p$ is a representation of the algebra $\mathcal{A}$. Moreover, $\pi $ is an injection. Indeed, let $r_a \in \mathcal{M}_0$ and $r_a = 0$. This means that $\pi(a) = 0$ for $\mu $-almost all $p \in E$ where $\mu $ is a measure on $E$. Consequently, for $\mu $-almost all $p \in E$ and all $\psi \in L^2(\Gamma^p)$ one has $\pi_p(a)\psi = 0$, i.e.,
\[
(\pi_p(a)\psi )(p_1, p) = \int_{E_{\pi_M(p)}}a(p_1,p_2)\psi(p_2,p)dp_2 = \int_G a(p_1, pg)\psi(pg,p)dg = 0.
\]
We have made the substitution $p_2 = pg$. Let $\psi(p_1, p) = \overline{a(p, p_1)}$. We have
\[
(\pi_p(a)\psi) (p,p) = \int_G a(p,pg)\overline{a(p,pg)}dg = \int_G |a(p,pg)|^2 dg = 0.
\]
This means that the support of $a_p := a(p,pg)$ is of zero measure and, consequently, the same is valid for $a$. But $a$ is of class $C^{\infty }$; therefore $a = 0$ which ends the proof.
\par
Since the mapping $a \mapsto r_a$ is an isomorphism of algebras, every derivation $u: \mathcal{A} \rightarrow \mathcal{A}$ defines the derivation $\tilde{u}: \mathcal{M}_0 \rightarrow \mathcal{M}_0$ by
\[
\tilde{u}(r_a)= r_{u(a)}.
\]
This allows us to define the Einstein operator $\tilde{\textbf{G}}:\tilde{V} \rightarrow \tilde{V}$, where $\tilde{V} \subset \mathrm{Der}(\mathcal{M}_0)$, by
\[
\tilde{\textbf{G}}(\tilde{u})(r_a) = r_{\textbf{G}(u)a}.
\]
This is valid only for $u \in V_1 \oplus V_2$ since, for $u \in V_3$, $\textbf{G}(u)$ is not defined. Therefore, if $u \in V_1 \oplus V_2$ is an eigenvector of the Einstein operator $\textbf{G}$ with the generalized eigenvalue $\tau $ then
\[
\tilde{\textbf{G}}(\tilde{u}) = r_{(\tau u)(a))} = \tau \cdot
r_{u(a)} = \tau \cdot \tilde{u}(r_a).
\]
\par
We thus can write the ``dynamical system'' for our model in the following form
\begin{equation}
\tilde{\textbf{G}}(\tilde{u}) = \tau \cdot \tilde{u}
\label{DS1}
\end{equation}%
for $u\in V_1 \oplus V_2$, and

\begin{equation}
\sigma _{t}^{\varphi }(r_{a}(p))=e^{itH_{p}^{\varphi
}}r_{a}(p)e^{-itH_{p}^{\varphi }}  \label{DS2}
\end{equation}
for every $p \in E$ and $a$ corresponding, in the unique manner, to
the inner derivation  ${ad}_a \in V_3$.
\par
Since our model consists of the differential algebra
$(\mathcal{A}, \mathrm{Der}(\mathcal{A}))$, its dynamical equation should
constrain both $\mathcal{A}$  and $\mathrm{Der}(\mathcal{A})$. This is
indeed the case: equation (\ref{DS1}) is essentially for derivations,
whereas equation (\ref{DS2}) is for the algebra.

\section{DYNAMICAL PROPERTIES OF THE QUAN\-TUM SECTOR}
The proposed model has a remarkable unifying power. In this section, we show how in its mathematical structure dynamics, probability and thermodynamics are unified.

\subsection{Noncommutative Probability}
In classical probability theory, basic objects of study are random
variables, i.e., measurable functions from a given probability space into
the set of reals $\mathbf{R}$ (equipped with the Borel $\sigma $-algebra
structure). With any such random variable $X$ there is associated a
probability measure $\mu _{X}(B)$ for any Borel set $B$. The measure $\mu
_{X}$ is also called the \textit{distribution\/} of $X$. In noncommutative
probability theory (Voiculescu, 1985; Voiculescu \textit{et al.}, 1992),
random variables are replaced by operators on a Hilbert space $\mathcal{H}$.
They are also called \textit{noncommutative random variables\/}. Instead of
working with the whole algebra $\mathcal{B}(\mathcal{H})$ of random
operators on $\mathcal{H}$ one usually restricts to a subalgebra which is a
von Neumann algebra. We recall that, by definition, it is a subalgebra $%
\mathcal{M}$ of $\mathcal{B}(\mathcal{H})$ containing the multiplicative
unit of $\mathcal{B}(\mathcal{H})$ and closed under the adjoint operation
and under taking limits in the weak topology on $\mathcal{B}(\mathcal{H})$,
i.e., topology induced by the linear functionals $b\mapsto \langle b\xi
,\eta \rangle ,\;\xi ,\eta \in \mathcal{H}$. Now, we must look for a
suitable counterpart of the probability measure on $\mathcal{M}$. We need
for it a kind of positivity and normalizability conditions. This is
implemented by the concept of state on the von Neumann algebra $\mathcal{M}$%
. A linear functional $\varphi :\mathcal{M}\rightarrow \mathbf{C}$ is a
\textit{state\/} on $\mathcal{M}$ if it takes nonnegative values on positive
operators on $\mathcal{M}$, and satisfies the condition $\varphi (1)=1$. The
pair $(\mathcal{M},\varphi )$, where $\mathcal{M}$ is a von Neumann algebra
and $\varphi $ a state on $\mathcal{M}$ is called a \textit{noncommutative
probability space\/}; $\varphi $ is called \textit{probability measure\/} on
$\mathcal{M}$. We shall additionally assume (as it is often done) that $%
\varphi $ is a faithful and normal state on $\mathcal{M}$. \textit{Faithful\/%
} means that $\varphi $ does not annihilate any nonzero positive element of $%
\mathcal{M}$ [i.e., $\varphi (r)=0$ implies $r=0$ for any positive element $%
r\in \mathcal{M}$]. \textit{Normal\/} means that if $r\in \mathcal{M}$ is
the supremum of a monotonically increasing net $\{r_{i}\}$ in the collection
of positive elements of $\mathcal{M}$ then $\varphi (r)=supp(r_{i})$. The
motivation for the above definition of noncommutative probability space
comes from the fact that if $\mathcal{M}$ is a commutative von Neumann
algebra, $\mathcal{M}$ is naturally isomorphic with the algebra of bounded
measurable functions (modulo equality almost everywhere) on an interval.

We thus have an en\-sem\-ble of non\-com\-muta\-ti\-ve prob\-a\-bil\-ity
spa\-ces $(\mathcal{M},\varphi )_{\varphi \in F}$ where $F$ denotes a
collection of normal and faithful states on $\mathcal{M}$. As we have seen
in the preceding subsection, each member $(\mathcal{M},\varphi )$ of this
ensemble is also a \textquotedblleft dynamic object\textquotedblright\
defining the modular evolution $\sigma _{s}^{\varphi }$. In this context it
seems natural that every noncommutative probability measure $\varphi $
determines its own dynamics of random operators [for more see\ (Heller
\textit{et al.}, 2005b)]. In this sense, two so far independent concepts are
unified: every dynamics is probabilistic and every probability is dynamic.

\subsection{Dynamics, Probability and Thermodynamics}

For the physicist any probabilistic dynamics is inseparably linked with
thermodynamic properties. To see that this is also the case in the
noncommutative context let us first remember some theoretical concepts.

A state $\varphi $ on the von Neumann algebra $\mathcal{M}$ is said to
satisfy the \textit{Kubo-Martin-Schwinger condition\/} (at inverse
temperature $\beta $), or is simply said to be a \textit{KMS state\/}, with
respect to a one-parameter group $\{\sigma _{s}:s\in \mathbf{R}\}$ of
automorphisms of $\mathcal{M}$ if, for each $A,B\in \mathcal{M}$, there
exists a bounded continuous function on the strip $\{z\in \mathbf{C}:0\leq
Imz\leq 1\}$, $F:\{z\in \mathbf{C}:0\leq Imz\leq 1\}\rightarrow \mathbf{C}$,
which is analytic in the interior of the strip and satisfies the following
conditions
\[
F(s+\beta i)=\varphi (A\sigma _{s}(B))
\]%
and
\[
F(s)=\varphi (\sigma _{s}(B)A),
\]%
for all $s$ in $\mathbf{R}$. Here $\beta =1/kT$ with $k$ being the Boltzmann
constant and $T$ absolute temperature (Sunder, 1987).

Let $\varphi $ be a normal and faithful state, and $\sigma _{t}^{\varphi
},t\in \mathbf{R}$, the modular group. Then $\varphi $ is a KMS state with
respect to $\sigma _{t}^{\varphi }$ and satisfies the condition $\varphi
\circ \sigma _{t}^{\varphi }=\varphi $. Moreover, such a modular group $%
\sigma _{t}^{\varphi }$ is uniquely determined (Connes and Rovelli, 1994,
section 2.5).

In quantum field theoretical statistical mechanics, KMS states are
interpreted as thermal equilibrium states (at inverse temperature $\beta $).
Rovelli argued that also in the relativistic context equilibrium states can
be characterized as faithful states on the algebra of observables whose
modular group is $\sigma _{s}^{\varphi }$ (Rovelli, 1993; Martinetti and
Rovelli, 2003). If we adopt this interpretation, we can claim that in the
noncommutative regime of our model dynamics, probability and at least some
aspects of thermodynamics are unified in the same mathematical structure.

\section{RANDOM DYNAMICS OF THE CLO\-SED FRIEDMAN UNIVERSE \label{Frieddyn}}

To illustrate the random behaviour in our model, let us return to the
noncommutative version of the closed Friedman universe. For simplicity we consider the two dimensional case $M = [0, T] \times S^1$. It is obvious that
the random evolution in this world model is expected to occur at its
earliest and latest phases in naigbourhoods of its initial and final
singularities.

The representation of the algebra $\mathcal{A}$, $\pi_p:\mathcal{A}%
\rightarrow \mathcal{B}(L^2(\Gamma^p))$, where $\Gamma^p=\{(\eta ,\chi
,\lambda_1,\lambda ):\lambda_ 1\in \mathbf{R}\}$ for $p = (\eta ,\chi
,\lambda )$), is now given by
\[
(\pi_p(a)\psi )(\lambda_1)=\int_{\mathbf{R}}a(\eta ,\chi
,\lambda_{1},\lambda_2)\psi (\eta ,\chi ,\lambda_2,\lambda )d\lambda_2,
\]
$a\in \mathcal{A},\,\psi\in L^2(\Gamma^p)$, and $\lambda$ is fixed.

The isomorphisms $I_p:L^2(\mathbf{R})\rightarrow L^2(\Gamma^ p)$ are given by

\[
(I_{p}(\psi _{0})(\eta ,\chi ,\lambda _{1},\lambda )=\psi _{0}(\lambda _{1})
\]%
for $\psi _{0}\in L^{2}(\mathbf{R})$. In this case, the regular
representation assumes the form
\[
(\tilde{\pi}_{p}(a)(\psi _{0})(\lambda _{1})=\int_{\mathbf{R}}a(\eta ,\chi
,\lambda _{1},\lambda _{2})\psi _{0}(\lambda _{2})d\lambda _{2}=\int_{%
\mathbf{R}}a_{\eta ,\chi }(\lambda _{1},\lambda _{2})\psi _{0}(\lambda
_{2})d\lambda _{2}.
\]%
The operator $\tilde{\pi}_{p}(a)$ is Hermitian if $a_{\eta ,\chi }(\lambda
_{2},\lambda _{1})=\overline{a_{\eta ,\chi }(\lambda _{1},\lambda _{2})},$ $%
\lambda _{1},\lambda _{2}\in \mathbf{R},$ for every $(\eta ,\chi )\in M$. We
have the norm $\mathrm{ess\,sup}(||\tilde{\pi}_{p}(a)||) < \infty $. Therefore, $(\tilde{%
\pi}_{p}(a))_{p\in E}$ are random operators. The algebra $\mathcal{M}_{0}$
of equivalence classes (modulo equality everywhere) of bounded random
operators is of the form $\mathcal{M}_{0}=\{E\ni p\mapsto \tilde{\pi}%
_{p}(a)\in \mathcal{B}(L^{2}(\mathbf{R})):a\in \mathcal{A}\}$. It can be
shown (Pysiak, 2006) that $\mathcal{M}_{0}$ generates the von Neumann
algebra
\[
\mathcal{M}\simeq L^{\infty }(M,\mathcal{B}(L^{2}(\mathbf{R})).
\]

Let $A=(\tilde{\pi}_{p}(a))_{p\in E}$. Let us also notice that it is enough
to define the states on $\mathcal{M}_{0}$. On the strength of proposition
B.1 of (Pysiak, 2006) such normal states are of the form
\begin{equation}
\varphi (A)=\int_{M\times \mathbf{R}\times \mathbf{R}}\mathrm{a}(\eta ,\chi
,\lambda _{1},\lambda _{2})\rho (\eta ,\chi ,\lambda _{1},\lambda _{2})d\eta
d\chi d\lambda _{1}d\lambda _{2}  \label{functional}
\end{equation}%
where $\rho $ is the density function. It must be nonnegative, Hermitian and integrable with the corresponding
integral equal to 1. To be faithful it must satisfy the
condition: $\rho (\eta ,\chi ,\lambda _{1},\lambda _{2})>0$ (modulo zero
measure subsets). Of course, there is one-to-one correspondence between $%
\varphi $ and $\rho $.

The important fact is that functional (\ref{functional}) is well defined
also in the presence of the initial and final singularities. In the closed
Friedman world model both these singularities are malicious, and
consequently the outer center $Z$ consists only of constant functions (Heller
and Sasin, 1995b, 1999; Heller \textit{et al.}, 2005a), but this fact has no
influence on the form of functional (\ref{functional}). Therefore, we can
say that $\varphi (A)$ does not \textquotedblleft feel\textquotedblright\
any singularity. Let us also notice that the functional $\varphi (A)$
prolongs well to $Z$; namely, if $f\in Z$, one has
\[
\varphi (A)=k\int_{M\times \mathbf{R}\times \mathbf{R}}\rho (\eta ,\chi
,\lambda _{1},\lambda _{2})d\eta d\chi d\lambda _{1}d\lambda _{2}=k
\]%
where $k$ is a constant value of $f$. This means that from the macroscopic
point of view $\varphi (A)$ is constant.
\par
On the strength of the Tomita-Takesaki theorem, the functional $\varphi $ (which is normal and faithful) determines a modular group $\sigma^{\phi }_t, \, t\in \textbf{R}$  of automorphisms of the von Neumann algebra $\cal{M}$. In terms of this group one can define the (state dependent) dynamics of random operators $(\tilde{\pi}_p(a))_{p \in \bar{E}},\, a \in \bar{\cal{A}}$. Since the functional $\varphi $ does not feel singularities, they are irrelevant for the dynamics of the Friedman model on its fundamental level. They appear only in the process of taking the ratio $\bar{M}=\bar{E}/G$ when space-time $M$ emerges out of the noncommutative regime; bars over $M$ and $E$ denote here suitable completions of $M$ and $E$, correspondingly [for the analysis of this process see (Heller and
Sasin, 1995b, 1999; Heller \textit{et al.}, 2003)].

\section{TRANSITION TO GENERAL RELATIVITY AND QUANTUM MECHANICS}

It is clear that to go from our model to general relativity one must
\textquotedblleft restrict\textquotedblright\ the algebra $\mathcal{A}%
=C^{\infty }(\Gamma ,\mathbf{C})$ to its \textquotedblleft outer
center\textquotedblright\ $Z=\pi _{M}^{\ast }(C^{\infty }(M))$ which, being
isomorphic to $C^{\infty }(M)$, naturally reproduces the usual spacetime
geometry. It is interesting that this can also be done with the help of the
following \textquotedblleft averaging\textquotedblright\ procedure. Let $%
\tilde{\mathcal{A}}$ be the extension of the algebra $\mathcal{A}$
\[
\mbox{$\tilde{\cal A}$}=\{a\in C^{\infty }(\Gamma ,\mathbf{C}):\forall x\in
M,a|\Gamma _{x}\in C_{c}^{\infty }(\Gamma _{x},\mathbf{C})\}
\]
where $\Gamma _{x}=E_{x}\times G$. Let further $\tilde{a}$ be the function
defined on $E\times G\times G$, corresponding to $a$, defined in the
following way: $\tilde{a}(p_{0},g_{1},g_{2})=a(p_{0}g_{1},g_{1}^{-1}g_{2})$.
Then the \textquotedblleft averaging\textquotedblright\ of $a$ is defined to
be
\begin{equation}
<a>=(\mathrm{Tr}a)(x)=\int_{G}\tilde{a}(p_{0},g,g)dg.  \label{averaging}
\end{equation}%
It is clear that $<a> \in Z$ which is isomorphic to the algebra $C^{\infty
}(M)$, and in terms of this algebra general relativity can be reconstructed
(Geroch 1972; Heller, 1992; Heller and Sasin 1995a).

The transition to quantum mechanics is even more interesting. If $a$ is a
Hermitian element of the algebra $\mathcal{A}$ then $ \pi_p(a)$ is a
Hermitian element of $(\mathcal{B}(\mathcal{H}^p))$ (since $\pi_p$ is a $*-$%
representation of the algebra $\mathcal{A}$). A random operator $%
r_a(p)=\pi_p(a)$ is Hermitian if $(r_a(p)\psi_{},\varphi )=(\psi
,r_a(p)\varphi )$. Moreover, it is a compact operator since $a$ has the
compact support. On the strength of the spectral theorem for Hermitian
compact operators in a separable Hilbert space, there exists in $\mathcal{H}%
^p$ an orthonormal countable Hilbert basis of eigenvectors $\{\psi_i\}_{i\in
I}$ of the Hermitian operator $r_a(p)$, and we can write its eigenvalue
equation as
\[
r_a(p)\psi_i(p)=\lambda_i(p)\psi_i(p)
\]
for every $p\in E$. Here $\lambda_i:E\rightarrow \mathbf{R}$ is a
generalized eigenvalue of the operator $r_a$. However, every measurement is
always done in a given local reference frame $p\in E$, and when such a
measurement has been done the generalized eigenvalue $\lambda_i$ collapses to the
eigenvalue $\lambda_i(p)$. Let us look deeper into the mechanism of this
collapse. Each act of measurement, performed at $p$, defines the isomorphism
$I_p ^{-1}: \mathcal{H}^p \rightarrow L^2(G)$ of Hilbert spaces (see,
subsection \ref{subs41}) which transfers the algebra of random operators
into the usual algebra of operators on the Hilbert space $L^2(G)$. In this
way, one obtains the usual quantum mechanics (on the group $G$).

For instance, let us apply the mapping $I_{p}^{-1}$ to the left hand side of
equation (\ref{TT2}), and the mapping $I_{p}$ to its right hand side. By
doing so, we obtain the usual Heisenberg equation for the evolution of $a\in
\mathcal{A}$
\[
\frac{d}{dt}\tilde{\pi}(a(t))=i[\tilde{H}^{\varphi },\tilde{\pi}(a(t))]
\]%
where $\tilde{\pi}(a)=I_{p}^{-1}\circ \pi _{p}\circ I_{p}$ and $\tilde{H}%
^{\varphi }=I_{p}^{-1}\circ H_{p}^{\varphi }\circ I_{p}$. The only
difference as compared with the usual Heisenberg equation is that the above equation
depends on the state $\varphi $. In more realistic models, to which the
Connes-Nikodym-Radon construction applies, even this difference will
disappear (Heller \textit{et al.}, 2005c).

In the light of the above analysis, the usual quantum mechanics is but a
theory of measurement within the larger structure of our model. When the act
of measurement is performed, the larger structure collapses to its
substructure known as quantum mechanics.

\section{PERSPECTIVES}

We do not claim that the model presented in this work should be regarded as
a concurrence with respect to theories like superstring theory or quantum
loop theory. First, it is not advanced enough and, second, we treat it
rather as a mean to deepen our understanding of conceptual subtleties that
are to be met along the road leading to the unification of physics. It is
not impossible that some elements of this model, or of its future more
mature forms, could be incorporated into better known approaches.

The noncommutative version of the closed Friedman world model, presented in
this work, is only a  ``test model'', but it
exhibits a remarkable property. Although in the original field equation
no matter term was explicitly included, the correct
components of the energy-momentum tensor (density and pressure) are obtained
as generalized eigenvalues of the Einstein operator. This effect can be regarded as essentially mitigating the strong dichotomy between geometry and matter inherent in the usual Einstein field equation.
\par
Another interesting problem related to the search for a fundamental theory
is whether the initial (or final) singularity will survive such a
revolution. Usually, either  ''yes'' or (more often)  ''no'' answers are given to
this question. Our model discloses the third possibility. Simple
calculations for a closed Friedman world model show that the random
character of dynamics on the fundamental level makes the question concerning
the singularities irrelevant. Singularities emerge from the noncommutative
regime together with the macroscopic spacetime. This result remains in
agreement with our previous works on classical singularities with the help
of noncommutative methods (Heller and Sasin, 1995b, 1999; Heller \textit{et
al.}, 2003).

Although the conceptual structure of our model seems esthetically satisfactory in many respects, we are aware of various its limitations; some of them could be overcome by enriching the architecture of the model. This could be done in many ways, perhaps the most obvious would be by taking into account the bialgebraic (or Hopf
algebraic) structure of the groupoid algebra [some coalgebra structure has been taken into
account in discussing observables of the model (Heller \textit{et al.}
2005a)]. The obvious next step to do is the elaboration of quantum field theoretical
aspects of the model (spinor bundles, Dirac's operator, etc.) together with
the gauge theoretic approach. Some preliminary work in this direction is
under way.

\bigskip

\noindent REFERENCES

\bigskip

Alicki, R. and Fannes, M. (2005). \textit{Reports on Mathematical Physics}
\textbf{55}, 47.

Carmeli, M. (1977). \textit{Group Theory and General Relativity}, McGraw-Hill, New York.

Chamseddine, A. H., Felder, G. and Fr\"{o}hlich, J. (1993). \textit{
Communications in Mathematical Physics } \textbf{155}, 205.

Chamseddine, A. H. (2001). \textit{Physical Letters} \textbf{B 504}, 33.

Chamseddine, E. H.(2004). \textit{Physical Review} \textbf{D 69}, 024015.

Connes, A. (1994). \textit{Noncommutative Geometry}, Academic Press, New
York.

Connes, A. and Rovelli, C. (1994). \textit{\ Classical and Quantum Gravity }
\textbf{11}, 2899.

Dixmier, J. (1981). \textit{Von Neumann Algebras}, North Holland, Amsterdam--New York--Oxford.

Dodson, C.T.J., (1978). \textit{International Journal of Theoretical Physics}
\textbf{17}, 389.

Dubois--Violette, M. (1988). \textit{Comptes Rendu de l' Academie des
Science, Paris.} \textbf{307}, 403.

Dubois-Violette, M. and Michor, P. W. (1994). \textit{Comptes Rendu de l'
Academie des Science, Paris. } \textbf{319}, 927.

Geroch, R. (1972). \textit{Communications in Mathematical Physics.} \textbf{26}, 271.

Heller, M. (1992). \textit{International Journal of Theoretical Physics} \textbf{31}, 272.

Heller, M. and Sasin, W. (1995a). \textit{International Journal of
Theoretical Physics} \textbf{34}, 387.

Heller, M. and Sasin, W. (1995b).  \textit{Journal of Mathematical Physics}
\textbf{36}, 3644.

Heller, M. and Sasin, W. (1999). \textit{General Relativity and Gravitation}
\textbf{31}, 555.

Heller, M., Sasin, W. and Lambert, D. (1997). \textit{Journal of
Mathematical Physics} \textbf{38}, 5840.

Heller, M., Sasin, W. and Odrzyg\'{o}\'{z}d\'{z}, Z. (1999). \textit{%
International Journal of Theoretical Physics} \textbf{38}, 1619.

Heller, M., Sasin, W. and Odrzyg\'{o}\'{z}d\'{z}, Z. (2000). \textit{Journal
of Mathematical Physics} \textbf{41}, 5168.

Heller, M., Odrzyg\'{o}\'{z}d\'{z}, Z., Pysiak, L. and Sasin, W. (2003).
\textit{International Journal of Theoretical Physics} \textbf{42}, 427.

Heller, M., Odrzyg\'{o}\'{z}d\'{z}, Z., Pysiak, L. and Sasin, W. (2004).
\textit{General Relativity and Gravitation} \textbf{36}, 111.

Heller, M., Odrzyg\'{o}\'{z}d\'{z}, Z. Pysiak, L. and Sasin, W. (2005a).
\textit{General Relativity and Gravitation} \textbf{37}, 541.

Heller, M., Pysiak, L. and Sasin, W. (2005b). \textit{International Journal
of Theoretical Physics} \textbf{44}, 619.

Heller, M., Pysiak, L. and Sasin, W. (2005c). \textit{Journal of Mathematical
Physics\/} \textbf{46}, 122501.

Madore, J. (1999). \textit{An Introduction to Noncommutative Differential
Geometry and Its Physical Applications}, 2nd ed. Cambridge University Press,
Cambridge.

Madore, J. and Mourad, J. (1998). \textit{Journal of Mathematical
Physics} \textbf{39}, 424.

Madore, J. and Saeger, L. A. (1998). \textit{Classical and Quantum Gravity}
\textbf{1998}, 811.

Martinetti, P. and Rovelli, C. (2003). \textit{Classical and Quantum Gravity}
\textbf{22}, 4919.

Misner, C. W., Thorne, K. S. and Wheeler, J. A. (1973), Gravitation,  Freeman and Comp.,San Francisco.

Moffat, J. W. (2000). \textit{Physics Letters} \textbf{B 491}, 345.

Murphy, G. J. (1990). \textit{$C^{\ast }$-algebras and Operator Theory\/},
Academic Press, New York-London.

Paterson, A. L. (1999). \textit{Groupoids, Inverse Semigroups and Their
Operator Algebras}, Birkh\"{a}user, Boston, Mass.

Pysiak, L. (2004). \textit{Demonstratio Mathematica} \textbf{37}, 661.

Pysiak, L. (2006). \textquotedblleft Time Flow in a Noncommutative
Regime\textquotedblright , \textit{International Journal of Theoretical
Physics\/}, in press.

Rovelli, C. (1993). \textit{Classical and Quantum Gravity} \textbf{10}, 1549.

Sitarz, A. (1994). \textit{Classical and Quantum Gravity} \textbf{11}, 2127.

Sunder, V. S. (1987). \textit{An Invitation to von Neumann Algebras},
Springer, New York--Berlin--Heidelberg.

Szabo, R. J. (2006). \textit{Classical and Quantum Gravity} \textbf{23}, R199.

Voiculecsu, D. V. (1985). ``Symmetries of some reduced free product $C^{\ast }$%
-algebras'', in: \textit{Operator Algebras and Their Connections with Topology
and Ergodic Theory}, Lecture Notes in Mathematics, vol. 1132. Springer,
Berlin--Heidelberg.

Voiculescu, D. V. Dykema, K. and Nica, A. (1992). \textit{Free Random
Variables}, CRM Monograph Series, No 1, American Mathematical Society,
Providence.

Wheeler, J. A. (1980). "Pregeometry: Motivations and Prospects", in:  \textit{Quantum Theory and Gravitation}, A. R. Marlow (ed.), Academic Press, New York.

\end{document}